\documentclass[twocolumn,showpacs,superscriptaddress,prl]{revtex4}
\usepackage{colordvi}
\usepackage{color}
\usepackage{epsfig}
\usepackage{graphicx}
\usepackage{dcolumn} 
\usepackage{amsmath} 

\begin{document}

\title{Nuclear stopping from $0.09A$ GeV
 to $1.93A$ GeV and its correlation to flow}

\date{ }
\author{W.~Reisdorf}
\affiliation{Gesellschaft f\"ur Schwerionenforschung, Darmstadt, Germany}
\author{A.~Andronic}    
\affiliation{Gesellschaft f\"ur Schwerionenforschung, Darmstadt, Germany}
\author{A.~Gobbi}     
\affiliation{Gesellschaft f\"ur Schwerionenforschung, Darmstadt, Germany}
\author{O.~N.~Hartmann}     
\affiliation{Gesellschaft f\"ur Schwerionenforschung, Darmstadt, Germany}
\author{N.~Herrmann}                
\affiliation{Physikalisches Institut der Universit\"at Heidelberg,
Heidelberg, Germany}   
\author{K.D.~Hildenbrand}     
\affiliation{Gesellschaft f\"ur Schwerionenforschung, Darmstadt, Germany}
\author{Y.J.~Kim}     
\affiliation{Gesellschaft f\"ur Schwerionenforschung, Darmstadt, Germany}
\affiliation{Korea University, Seoul, South Korea}  
\author{M.~Kirejczyk}     
\affiliation{Gesellschaft f\"ur Schwerionenforschung, Darmstadt, Germany}
\affiliation{Institute of Experimental Physics, Warsaw University, Poland}  
\author{P.~Koczo\'{n}}     
\affiliation{Gesellschaft f\"ur Schwerionenforschung, Darmstadt, Germany}
\author{T.~Kress}     
\affiliation{Gesellschaft f\"ur Schwerionenforschung, Darmstadt, Germany}
\author{Y.~Leifels}     
\affiliation{Gesellschaft f\"ur Schwerionenforschung, Darmstadt, Germany}
\author{A.~Sch\"{u}ttauf}     
\affiliation{Gesellschaft f\"ur Schwerionenforschung, Darmstadt, Germany}
\author{Z.~Tymi\'{n}ski}     
\affiliation{Gesellschaft f\"ur Schwerionenforschung, Darmstadt, Germany}
\affiliation{Institute of Experimental Physics, Warsaw University, Poland}  
\author{Z.G.~Xiao}           
\affiliation{Gesellschaft f\"ur Schwerionenforschung, Darmstadt, Germany}
\author{J.P.\,Alard}    
\affiliation{Laboratoire de Physique Corpusculaire, IN2P3/CNRS,
and Universit\'{e} Blaise Pascal, Clermont-Ferrand, France} 
\author{V.~Barret}     
\affiliation{Laboratoire de Physique Corpusculaire, IN2P3/CNRS,
and Universit\'{e} Blaise Pascal, Clermont-Ferrand, France} 
\author{Z.\,Basrak}       
\affiliation{Rudjer Boskovic Institute, Zagreb, Croatia}
\author{N.\,Bastid}     
\affiliation{Laboratoire de Physique Corpusculaire, IN2P3/CNRS,
and Universit\'{e} Blaise Pascal, Clermont-Ferrand, France} 
\author{M.L.\,Benabderrahmane}           
\affiliation{Physikalisches Institut der Universit\"at Heidelberg,
Heidelberg, Germany}   
\author{R.\,\v{C}aplar}          
\affiliation{Rudjer Boskovic Institute, Zagreb, Croatia}
\author{P.~Crochet}       
\affiliation{Laboratoire de Physique Corpusculaire, IN2P3/CNRS,
and Universit\'{e} Blaise Pascal, Clermont-Ferrand, France} 
\author{P.\,Dupieux}         
\affiliation{Laboratoire de Physique Corpusculaire, IN2P3/CNRS,
and Universit\'{e} Blaise Pascal, Clermont-Ferrand, France} 
\author{M.\,D\v{z}elalija}           
\affiliation{Rudjer Boskovic Institute, Zagreb, Croatia}
\author{Z.\,Fodor}            
\affiliation{Central Research Institute for Physics, Budapest, Hungary} 
\author{Y.\,Grishkin}        
\affiliation{Institute for Theoretical and Experimental Physics, Moscow, Russia}
\author{B.\,Hong}         
\affiliation{Korea University, Seoul, South Korea}  
\author{J.\,Kecskemeti}             
\affiliation{Central Research Institute for Physics, Budapest, Hungary} 
\author{M.\,Korolija}           
\affiliation{Rudjer Boskovic Institute, Zagreb, Croatia}
\author{R.\,Kotte}          
\affiliation{Forschungszentrum Rossendorf, Dresden, Germany}   
\author{A.\,Lebedev}         
\affiliation{Institute for Theoretical and Experimental Physics, Moscow, Russia}
\author{X.~Lopez}        
\affiliation{Laboratoire de Physique Corpusculaire, IN2P3/CNRS,
and Universit\'{e} Blaise Pascal, Clermont-Ferrand, France} 
\author{M.\,Merschmeyer}               
\affiliation{Physikalisches Institut der Universit\"at Heidelberg,
Heidelberg, Germany}   
\author{J.\,M\"{o}sner}       
\affiliation{Forschungszentrum Rossendorf, Dresden, Germany}   
\author{W.\,Neubert}          
\affiliation{Forschungszentrum Rossendorf, Dresden, Germany}   
\author{D.\,Pelte}               
\affiliation{Physikalisches Institut der Universit\"at Heidelberg,
Heidelberg, Germany}   
\author{M.\,Petrovici}           
\affiliation{National Institute for Nuclear Physics and Engineering, Bucharest,
Romania}
\author{F.\,Rami}        
\affiliation{Institut de Recherches Subatomiques, IN2P3-CNRS, Universit\'e
Louis Pasteur, Strasbourg, France}
\author{B.\,de Schauenburg}          
\affiliation{Institut de Recherches Subatomiques, IN2P3-CNRS, Universit\'e
Louis Pasteur, Strasbourg, France}
\author{Z.\,Seres}              
\affiliation{Central Research Institute for Physics, Budapest, Hungary} 
\author{B.\,Sikora}             
\affiliation{Institute of Experimental Physics, Warsaw University, Poland}  
\author{K.S.\,Sim}       
\affiliation{Korea University, Seoul, South Korea}  
\author{V.\,Simion}            
\affiliation{National Institute for Nuclear Physics and Engineering, Bucharest,
Romania}
\author{K.\,Siwek-Wilczy\'nska}          
\affiliation{Institute of Experimental Physics, Warsaw University, Poland}  
\author{V.\,Smolyankin}         
\affiliation{Institute for Theoretical and Experimental Physics, Moscow, Russia}
\author{M.\,Stockmeier}              
\affiliation{Physikalisches Institut der Universit\"at Heidelberg,
Heidelberg, Germany}   
\author{G.\,Stoicea}                  
\affiliation{National Institute for Nuclear Physics and Engineering, Bucharest,
Romania}
\author{P.\,Wagner}         
\affiliation{Institut de Recherches Subatomiques, IN2P3-CNRS, Universit\'e
Louis Pasteur, Strasbourg, France}
\author{K.~Wi\'{s}niewski}             
\affiliation{Institute of Experimental Physics, Warsaw University, Poland}  
\author{D.\,Wohlfarth}           
\affiliation{Forschungszentrum Rossendorf, Dresden, Germany}   
\author{I.\,Yushmanov}     
\affiliation{Kurchatov Institute, Moscow, Russia}
\author{A.\,Zhilin}            
\affiliation{Institute for Theoretical and Experimental Physics, Moscow, Russia}
%-------------------------------------------------------------------
\collaboration{FOPI Collaboration}
\noaffiliation

%-----------------------------------------------------------------------

\begin{abstract}
We present a complete systematics (excitation functions and system-size
dependences)
of global stopping and sideflow 
for heavy ion reactions in the energy range between $0.09A$ GeV and $1.93A$ GeV.
For the heaviest system, Au+Au, we observe a plateau of maximal stopping
extending from about $0.2A$ to $0.8A$ GeV with a fast drop on both sides.
The degree of stopping, which is shown to remain
significantly below the expectations of a full
stopping scenario, is found to be highly correlated to the amount of sideflow.
\end{abstract}
%-----------------------------------------------------------------------


\maketitle
%-----------------------------------------------------------------------

In the last two decades collisions between accelerated heavy ions
have become a major laboratory tool in large scale efforts to learn more about
the properties of hot and compressed nuclear matter.
Nuclei are small objects however:
from the Bethe-Weizs\"{a}cker formula describing
empirical nuclear masses it follows that surface and Coulomb effects,
that are not present in the interior of macroscopic nuclear objects, such as
neutron stars, reduce the bulk binding energy per nucleon of 16 MeV to about
half the value, i.e. 8 MeV.
For this reason the one-fluid hydrodynamics approaches~\cite{scheid74,
stoecker86,clare86} that were favoured in the early days to link experimental
observations and fundamental properties of nuclear matter, such as the relation
between pressure, density and temperature (Equation of State, EoS) were
not trivially justified. 
The first convincing evidence of
collective (side) flow~\cite{gustafsson84,renfordt84}
raised hopes that this connection could be established
successfully~\cite{buchwald84}.
Meanwhile, flow phenomena in heavy ion reactions have been studied 
extensively~\cite{reisdorf97}.
With the advent in the mid eighties of event simulation codes implementing
microscopic transport theory~\cite{bertsch88} it soon became clear that 
non-equilibrium effects in such reactions could not be ignored.
In one of the most recent attempts~\cite{danielewicz02} to derive constraints on
the EoS from heavy ion data (which also mentions some of the earlier works)
it is made clear that progress on this topic requires improved understanding of
the momentum dependence of the mean fields generated in a heavy ion
collision (first introduced in refs.~\cite{aichelin87,gale87})
as well as an extensive adjustment to experimental information on the degree of
stopping achieved.

This Letter presents a complete systematics of stopping properties in
heavy ion collisions at SIS energies. 
Our data show  that  the degree of stopping achieved
in a heavy ion collision 
1) reaches a well defined plateau of maximal stopping centered around 
 $0.5A \pm 0.3A$ GeV with a fast drop on both edges,
2) stays significantly below the hydrodynamic limit at all energies,
3) shows a strong correlation to the global sideflow suggesting that
parametric adjustments to both observables might require an iterative
procedure.

The experiments were performed at the heavy ion accelerator SIS of
GSI/Darmstadt using the large acceptance FOPI detector \cite{ritman95}.
A total of 25 system-energies are analysed for this work (energies in $A$ GeV
are given in parentheses):
$^{40}$Ca+$^{40}$Ca (0.4, 0.6, 0.8, 1.0, 1.5, 1.93),
$^{58}$Ni+$^{58}$Ni (0.09, 0.15, 0.25, 0.4),            
$^{96}$Ru+$^{96}$Ru (0.4, 1.5),                    
$^{129}$Xe+$^{133}$Cs$^{127}$I (0.15, 0.25, 0.4),             
$^{197}$Au+$^{197}$Au (0.09, 0.12, 0.15, 0.25, 0.4, 0.6, 0.8, 1.0, 1.2, 1.5).
Particle tracking  and energy loss determinations are
done using two drift chambers, the CDC (covering polar angles
between $35^\circ$ and $135^\circ$) and 
the Helitron ($9^\circ-26^\circ$), both
located inside a superconducting solenoid operated at a magnetic field of
0.6T.
A set of scintillator arrays, Plastic Wall $(7^\circ-30^\circ)$, 
Zero Degree Detector
$(1.4^\circ-7^\circ)$, and Barrel $(42^\circ-120^\circ)$,
 allow us to measure the time of flight
and, below $30^\circ$, also the energy loss.
For the low-energy experiments (up to $0.4A$ GeV) the Helitron was replaced by
an array of gas ionization chambers~\cite{gobbi93} allowing charge
identification of heavier clusters.
The velocity resolution below $30^\circ$ was $(0.5-1.5)\%$,
the momentum resolution in the
CDC was $(4-12)\%$ for momenta of 0.5 to 2 GeV/c, respectively.
Use of CDC and Helitron allows the identification of pions, as well as
good isotope separation for  hydrogen and
helium clusters in a large part of momentum space.
Heavier clusters are separated  by nuclear charge.

Collision centrality selection was obtained by binning 
distributions of either the detected
charge particle multiplicity, MUL, or the ratio of total transverse to
longitudinal kinetic energies in the center-of-mass (c.o.m) system, $Erat$.

\begin{figure}[t]
\centering{\mbox{\hspace{-1.0cm} 
\epsfig{file=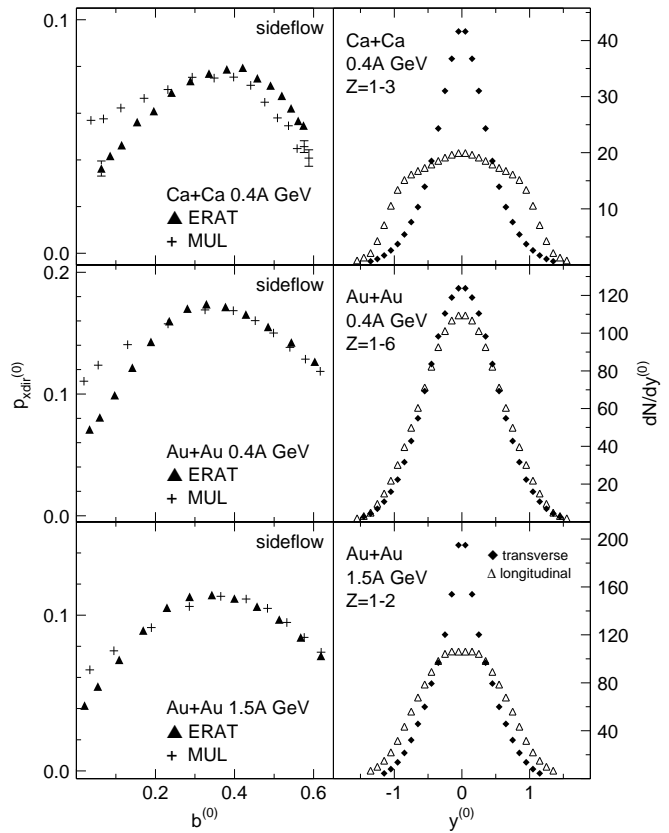,width=10.5cm}}}
\vspace{-0.5cm}
%
% Fig. 1
\caption{
Left panels: global sideflow as a function of the scaled impact parameter
(triangles: $Erat$ selection, crosses: multiplicity selection).
Right panels: global scaled transverse and longitudinal rapidity distributions
for the $Erat$ selection $b^{(0)}<0.15$. Data for three indicated 
system-energies are shown.
Systematic errors are discussed in the text.
}
\label{f:fig1}
\end{figure}

The scaled directed flow is $p_{xdir}^{(0)} \equiv p_{xdir}/u_{1cm}$ where
$p_{xdir}=\sum sign(y) Z u_x/\sum Z$ ($Z$~~fragment charge, $u_{1cm}$ 
spatial part of the c.o.m.
projectile 4-velocity, $u_x \equiv \beta_x\gamma$ 
projection of the fragment 4-velocity on the
reaction plane~\cite{danody}). The sum is over all measured charged particles
with $Z < 10$, excluding pions, and
$y$ is the c.o.m. rapidity. 
In the left panels of Fig.~\ref{f:fig1} we show the scaled impact
parameter, $b^{(0)}=b/b_{max}$, dependence of $p_{xdir}^{(0)}$
for three system-energies,
$^{40}$Ca + $^{40}$Ca at $0.4A$ GeV, and $^{197}$Au+$^{197}$Au ($0.4A$ and
$1.5A$ GeV).
We take $b_{max} = 1.15 (A_{P}^{1/3} + A_{T}^{1/3})$~fm and estimate $b$
from the measured differential cross sections for the $Erat$ or the
multiplicity distributions, using a geometrical sharp-cut approximation.
The $Erat$ selections show better impact parameter resolution for the most
central collisions than the multiplicity selections.
The maximum value of $p_{xdir}^{(0)}$ is a robust observable 
that does not depend
significantly on both the selection method and the apparatus filter.
For a comparative study of alternate, more differential measures of sideflow
in the published literature we refer to the reviews~\cite{reisdorf97}.

%\vspace{1.0cm}
To obtain information on the degree of stopping, we measure rapidity
distributions of the emitted particles.
In an XYZ Cartesian coordinate system with the Z-axis oriented in the beam
direction and the X-axis in the reaction plane, we define
the (conventional) longitudinal rapidity distribution in the
beam direction, and 
 the transverse rapidity averaged over the X and
Y directions by projecting on a transverse laboratory fixed axis. 
All rapidities are in the c.o.m. and are scaled by the beam rapidity.
Results for the same system-energies as in the left panels are shown in
the right panels of Fig.~\ref{f:fig1}.

The distributions are {\em full event} observables: all emitted charged
baryonic particles are added, weighted with their respective nuclear charges.
The total charge accounted for represents typically $95\%$ of the system charge
or more, except for data below 150A MeV, where threshold effects prevent 
complete coverage of fragments with charge $Z>6$.
Where necessary, bi-dimensional fitting procedures 
in $u_t$ vs $y$ space, which require reflection symmetry, have been used
\cite{reisdorf} to interpolate and partially extrapolate the measured spectra
($u_t \equiv \beta _t\gamma$ transverse 4-velocity, $y \equiv y_z$ longitudinal
rapidity).
Extrapolation corrections to obtain $4\pi$ -distributions     were less than
$10\%$ for protons and clusters with nuclear charge $Z=1, 2$  and 
up to a maximum of $30\%$ for heavier fragments.

As a measure of the degree of stopping we present data for an observable dubbed
{\em vartl}, the ratio of the variances of the transverse to that of the
longitudinal  rapidity distributions.
For an isotropic thermal source $vartl$ should be equal to unity.
Flow and transparency may change {\em vartl} however.
Early, ideal hydrodynamics estimates~\cite{scheid74} 
predict in this energy regime the phenomenon of {\em squeeze-out}, resulting,
for {\em central} collisions, in enhanced emission at polar (c.o.m.) angles of
$90^{\circ}$: {\em vartl} then becomes significantly larger than unity.
To avoid the uncertainties and distortions of the measured far-out tails of
these distributions, we limit the integration to the interval between +1 and
-1 in the scaled values.
By definition $Erat$ and $vartl$ are related, although technically different.
Autocorrelations in $vartl$, as well as $p_{xdir}$, were avoided by 
removing the particle of interest from the selection
criterion $Erat$ or, respectively, the transverse-momentum
vector~\cite{danody} determining
the reaction plane.

The observable {\em vartl} is also related in spirit to
$R = (2/\pi)\sum |p_{ti}| / \sum |p_{zi}|$ suggested in ref.~\cite{stroebele83}
and used in earlier stopping studies~\cite{gutbrod89}.
Longitudinal rapidity distributions for selected particles,
measured with the PLASTIC BALL
\cite{gutbrod89} were published~\cite{gutbrod90}
for a limited set of systems and beam
energies.

\begin{figure}[t]
\centering{\mbox{\hspace{-1.5cm} 
\epsfig{file=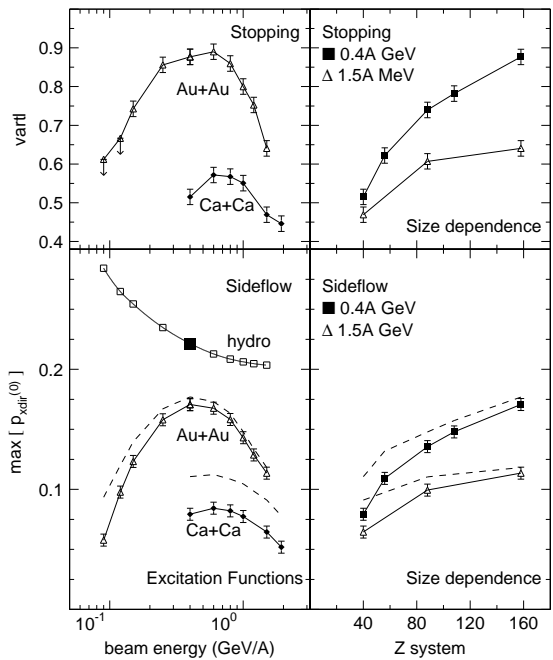,width=8cm,clip}}}
\vspace{-0.5cm}
% Fig.2
\caption{
Excitation functions (left) and size dependences (right) of the   
degree of stopping (upper panels) and maximal global sideflow (lower
panels). The dashed curves result from a correction for statistical       
fluctuations in the reaction plane orientation.
The curve labelled 'hydro' is an estimate of the hydrodynamic limit
for the sideflow. See text.
}
\label{f:syst}
\end{figure}

At this point we note that all statistical errors in this work are
significantly smaller than the systematic errors.
Estimates of the latter, varying between 5 and 12~$\%$ for the rapidity
distributions, are based on comparisons of measured yields in the c.o.m.
backward/forward hemispheres, checks of the overall charge balance and
technical variations of the fitting procedure.
For $p_{xdir}^{(0)}$ ($vartl$) a constant uncertainty of 0.008 (0.04)
describes best
the overall error. Point-by-point fluctuations (shown in Fig.~\ref{f:syst})
are smaller: 0.005 (0.02).
Simulations with the transport code IQMD (based on Quantum Molecular
Dynamics~\cite{hartnack}), together with detector filters, 
were also used to assess these errors.
 
The upper-left panel of Fig.~2 shows excitation functions of {\em vartl} for
Au+Au and Ca+Ca  in central collisions corresponding to scaled  impact
parameters $b^{(0)} < 0.15$. 
Interesting aspects of these data are
a) the occurrence, for Au+Au, of a plateau of 
maximal stopping extending from $0.2A$ GeV to $0.8A$ GeV with a fast
drop on both sides,
b) the fact that
{\em vartl}  never reaches values above or even close to unity,
and
c) the large system-size dependence evidenced by comparing the
Au+Au with the Ca+Ca data.

The smaller values for the lighter system offer   a key
argument in favour of interpreting {\em vartl} $< 1$
as evidence for partial transparency, rather than the dominance of longitudinal
over transversal pressure gradients.     

More detailed information on the system-size dependence at two incident
beam energies ($0.4A$ GeV and $1.5A$ GeV) is shown in the upper right panel of
Fig.2 which also demonstrates that, especially at the lower energy, there is no
evident saturation as one goes from Ca+Ca, via Ni+Ni, Ru+Ru, Xe+CsI all the way
to Au+Au.

The two lower panels of the figure show by comparison with the two upper panels 
an impressive correlation with the maximum (see Fig.~\ref{f:fig1})
scaled directed
sideflow, max~$[p_{xdir}^{(0)}]$, both in the excitation functions
and in the size dependence.
In particular sideflow is
{\em positively} correlated with {\em vartl} 
as shown in Fig.~3 for  22 system-energies (data for $E/A < 0.15$GeV
were omitted
because only upper limits on $vartl$ could be determined).                  

\begin{figure}[b]
\centering{\mbox{\hspace{-1.5cm} 
\epsfig{file=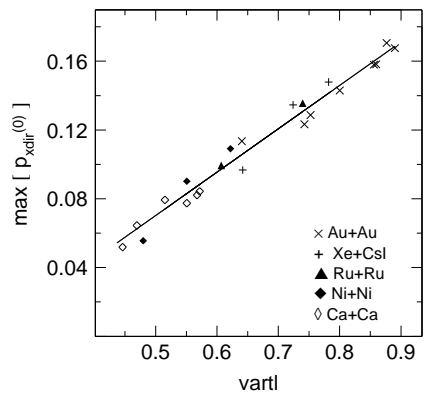,width=6.5cm}}}
\vspace{-0.5cm}
%
% Fig.3
\caption{
Correlation between the maximal directed sideflow and the
degree of stopping. The line is a linear 
least squares fit to the data, which extend from 0.15 to 1.93 $A$ GeV.
}
\label{f:vartl}
\end{figure}

Although the two global observables are measured at different average
impact parameters,
the most natural explanation of the observed correlation is that sideflow 
is resulting
from the pressure gradient between  participant and spectator matter,
which in turn grows with the degree of stopping.
Note that properly scaled sideflow is essential to reveal this
correlation.
Both observables are also expected to be influenced by the statistics of
elementary collisions.
The estimated~\cite{ollitrault} effect on sideflow is shown in Fig.2.

Using the Rankine-Hugoniot-Taub one-dimensional shock equations
together with a realistic equation-of-state 
(SLy230a from ref.~\cite{chabanat}), we have estimated the maximal densities
and pressures  $\mathcal{P}$ to be expected under the full stopping assumption
for the energy regime of interest here.
Characterizing the system size by $R_{sys}$ and the (passing)
time by $t_{pass}$, the scaled global sideflow should behave as 
$\mathit{grad}\mathcal{P} \times t_{pass}/u_{1cm} \approx 
(\mathcal{P}/R_{sys}) (R_{sys}/u_{1cm})/u_{1cm} = \mathcal{P}/u_{1cm}^2$.
The latter quantity, with a normalization to be discussed below, is
plotted in Fig.~\ref{f:syst} (lower left panel)
and compared to the measured scaled sideflow.
Note that ideal-gas one-fluid hydrodynamics predicts~\cite{balazs}
a size independent flat excitation function  
($R_{sys}$ drops out, and $\mathcal{P}$ rises linearly with
$u_{1cm}^2$).

Our estimate of the shape of the excitation function
does not yield the absolute value of max~$[p_{xdir}^{(0)}]$,
which depends on the geometrical details of the collision.
We opted for an estimate of the normalization by
performing an IQMD calculation using, instead of the free-space nucleon-nucleon
elastic cross sections, values enhanced by a factor two over the known
free space values, thus mocking up approximately the full stopping situation
(we found $vartl = 1.23$, confirming 
qualitatively early predictions~\cite{scheid74}).
The resulting maximal sideflow value (0.225 at $0.4A$ GeV) is marked by
a full square in the
Figure . This serves  to give an idea on what
to expect in the ideal hydrodynamics (non-viscous, one-fluid) limit.
Introducing still higher elementary cross sections in the simulation does not
change the prediction of $vartl$ significantly 
as expected in the hydrodynamic limit
and confirming the relevance of $vartl$ to assess stopping.

The difference
between the theoretical estimate and the experimental observation is likely to
be the finite viscosity of hot nuclear matter, or, equivalently, the finite
ratio $R_{sys}/\lambda_f$ (which is the Reynolds number with $\lambda_f$ the
mean free path), which is not sufficiently large to produce fully macroscopic
ideal hydrodynamic behaviour in such collisions.
The significant difference of typically $30\%$ in the plateau region,
together with the observed strong system-size dependence
(already known from earlier studies~\cite{gutbrod89,reisdorf97} 
of the flow of selected ejectiles),
stress the 'small-object' character of nuclei:
finite-size (surface) effects are large.
The most interesting, and new, aspect of the full-event
data, however, is the rapid drop
away from the plateau.
In particular, the gradual rise of the theoretical prediction at the low energy
end is totally suppressed by both the onset of partially attractive mean fields
and the increase of Pauli blocking below 0.4A GeV \cite{daniel}.
On the high energy end (1.5A GeV) of our data, looking again at 
Fig.~\ref{f:syst},
one can estimate that the effective average pressures achieved
are as much as a factor
two lower than the full-stop scenario would predict. 
In-medium modifications of elementary cross sections \cite{li93}, caused
possibly by  nucleon-mass changes at high density, might be needed
to account for the
apparent rise of transparency in an energy regime where strong effects
of Pauli blocking are no longer expected. 

Our observation of partial transparency confirms qualitatively earlier  
isospin tracer experiments~\cite{hong} and,
together with its close correlation to sideflow, has important consequences 
for the ongoing efforts to extract the Equation of State from heavy ion data
in the non-equilibrium situation~\cite{gaitanos}, 
as well as for attempts to understand in-medium effects on mesonic and baryonic
particles, since the effective densities in the early phase of the reactions
are expected to be strongly correlated to the degree of stopping.  

This work was partly supported by the German BMBF under contract 06HD953 and
RUM-99/010 and by KRF under Grant 2002-015-CS0009.

%---------------------------------------------------------------------
%  #ref

%---------------------------------------------------------------------
% #fig
\end{document}